\documentclass[12pt]{article}
\usepackage[latin1]{inputenc}
\usepackage{amsmath}
\usepackage{latexsym}
\usepackage[round]{natbib}

\usepackage[dvips,final]{graphicx}
\usepackage{epsfig}

\newcommand{\D}{{\rm D}}

\newcommand{\mi}{\mathrm{i}}

\begin{document}

\title{On the sensitivity of coastal quasigeostrophic edge wave interaction to bottom boundary characteristics: \\  possible implications for eddy parameterizations}
\author{Tapani Stipa \\ Finnish Institute of Marine Research \\ PO Box 33 \\ FIN-00931 Helsinki \\ email Tapani.Stipa@fimr.fi \\ tel. +358-9-613941}
\date{\texttt{$Revision: 1.20 $ $Date: 2003/01/31 07:57:37 $}}

\maketitle

\begin{abstract}

The Eady problem of baroclinic instability as applicable to quasi-geostrophic oceanic flows with zero internal PV gradients is revisited by introducing a mild slope and Ekman pumping on the lower boundary. The solution behaviour is determined by the isopycnal slope relative to either the bottom slope or the ratio of Ekman depth to horizontal wavenumber. Attention is paid to the physical interpretation of the growing, decaying and stable disturbances, with emphasis on the intimate connection between the quasigeostrophic edge waves and Eady waves, and the role of the isopycnal slope for the stability properties as opposed to the bottom density gradient. The disturbance structure is found to be strongly influenced by the boundary conditions.

For a sloping bottom boundary, the growth rate is enhanced for the most unstable waves if the isopycnals tilt in the same direction as the bottom, but in general non-standard boundary conditions tend to retard the growth of disturbances. In particular, the existence of the long- and short-wave cutoffs is found to be very sensitive to boundary conditions, both for the sloping topography and the Ekman pumping. It is suggested that any cutoffs for the growth rate in an Eady-like problem actually result from the chosen boundary conditions. However, for a certain range of parameters, the maximum growth rate is comparable to that found in the original Eady problem, which may explain the fair success enjoyed by recent eddy parameterizations basing their timescale on the Eady growth rate.


\end{abstract}

\section{Introduction}
\label{sec:intro}

The importance of oceanic transports induced by mesoscale eddies has been demonstrated in the previous decade for passive tracers  \citep{Danabasoglu.etal.Science.1994} as well as, tentatively, for ecological, reactive tracers \citep{Oschlies.Garcon.1998.Nature}. The formation of these mid-ocean eddy transports is frequently described with the baroclinic, quasigeostrophic stability problem of \citet{Eady.1949} as a prototype.

Less attention has been devoted to eddy transports in the coastal areas. As opposed to the original problem formulated by Eady with only horizontal boundaries, coastal areas typically have a sloping bottom and, possibly as a consequence,  buoyant coastal currents may persist over long distances under apparently unstable conditions. The stability of such currents is a difficult problem that hitherto has defied unapproximated analysis; but see \cite{Poulin.Swaters.1999} for recent progress. Extensions of the tractable analysis due to Eady are pursued in this study as a source of guidance for understanding the instability and in particular, the consequent eddy fluxes associated with coastal currents.

In an atmospheric setting, the effect of a small bottom slope in the Eady problem was studied by \citet{Blumsack.Gierasch.1972} as a prototype for Martian climate. \citet{Mechoso.1980,Mechoso.1980.antarctica} extended the theory to deal with two sloping boundaries and applied it to the terrestrial atmosphere, with an oceanographic discussion in \citet{Mechoso.Sinton.1981}. \citet{deSzoeke.1975} developed the bottom profile into a perturbation series and discussed the resulting hybrid instabilities.

These studies found substantially modified stability properties for changes of the bottom slope, and as demonstrated at the end of the present investigation, it is possible to relax the necessary condition for instability altogether to make the flow unconditionally stable. For the eddy parameterizations presently used in numerical models of the ocean,  the growth rate is determined from the original Eady theory. Hence, such parameterisations do not appear to be immediately applicable to coastal situations with sloping boundaries. 

The modifications caused by Ekman layers at the bottom boundary layer are not obvious either. \citet{Holopainen.1961.friction} discusses the effect of Ekman pumping in a two-layer model on a $\beta$-plane, noting that friction may broaden the unstable wave-band while still retaining the short- and long-wave cutoffs. \citet{Williams.Robinson.1974.geneady} generalized the Eady problem by combining the effects of a varying stratification with Ekman pumping at one or two boundaries. These authors found a destabilizing effect on the short waves, i.e.~there is no parameter range with absolute stability. 

The instability in the Eady problem is often discussed as resulting from the meridional density- or temperature-gradient at the boundary. While this argument is mathematically correct, the actual reason for the boundary gradient is the intersection of sloping interior isopycnals with a horizontal boundary. The intimate connection between  the topographically modified Eady waves and  the quasigeostrophic edge waves touched upon by \citet{Rhines.edge.1970} is in the present context found to be important for the stability properties.

In the present work the growth rate and structure of the disturbances in baroclinic instability of the Eady problem in the presence of topography and friction are studied in more detail than previously (cf.\ \citet{Blumsack.Gierasch.1972}, \citet{Williams.Robinson.1974.geneady}), emphasizing the importance of the isopycnal slope for the physical interpretation. The lower boundary is modified to include a slope (Section \ref{sec:fullsol}) and Ekman pumping (Section \ref{sec:eady-problem-with-Ekman}) in order to clarify their effects on the oceanic stability problem. 

The structure and stability properties of the solutions are found to be substantially modified by the boundary conditions and are discussed in Section \ref{sec:discuss} with emphasis on the physical background. From the analysis the pure Eady problem emerges  as a special case of a more general baroclinic instability problem.

\section{The Eady problem with boundary layer slope}
\label{sec:fullsol}

\begin{figure}[tp]
  \begin{center}
    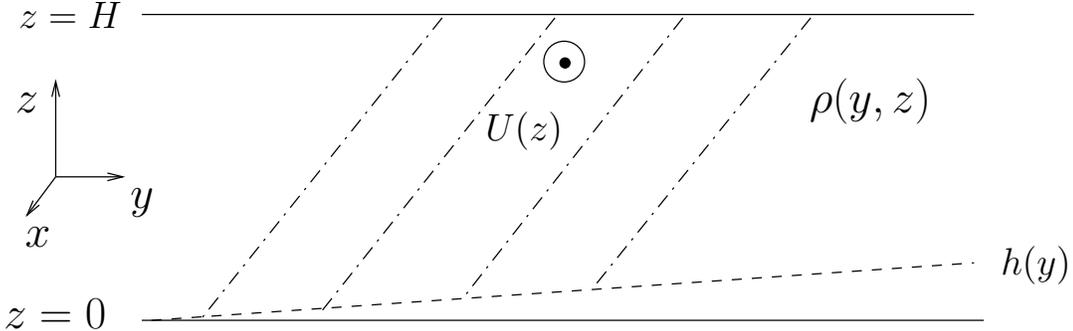
    \caption{The Eady problem as applied to a buoyant flow with an infinitesimal bottom slope. The isopycnals of the basic-state density field $\rho (y,z)$ slope with a constant angle throughout the fluid interior. Rigid lids, one horizontal on the top and one sloping on the bottom at $h(y) \ll  H$, prevent vertical movements across them. The flow $U(z)$ has constant shear, with $U(z=0)=0$.}
    \label{fig:eadyfig.z}
  \end{center}
\end{figure}

The configuration characterizing the continuous stability problem formulated by \citet{Eady.1949} with a sloping lower boundary added is depicted in Figure \ref{fig:eadyfig.z}. For the atmosphere, an analytical approach to the problem has been  pioneered by \citet{Blumsack.Gierasch.1972}, whose analysis is extended here. Dimensional forms of the equations are retained in order to keep close track of the physics. The relevant scaling arguments can be found in e.g. \citet{Gill.1982}, whose notation we follow; a brief derivation of the governing equations is presented to assist the discussion in Section \ref{sec:discuss}.

The conservation equation for pseudopotential vorticity $q$ in an incompressible flow reads
  \begin{equation}
    \label{eq:qgpveq}
    \frac{\D }{\D t} \left( \frac{\partial^2 p}{\partial x^2} + \frac{\partial^2 p}{\partial y^2} + \frac{\partial }{\partial z} \left( \frac{f^2 }{N^2} \frac{\partial p}{\partial z} \right) \right) = \frac{\D }{\D t} q =0.
  \end{equation}
Here $\frac{D}{D t}=\frac{\partial }{\partial t} + U \frac{\partial }{\partial x} + v \frac{\partial }{\partial y}$ is the geostrophic advection operator with $ U = ( -f \rho_0 )^{-1} \partial p / \partial y $  as the basic-state flow in the positive $x$-direction, $v = ( f \rho_0 )^{-1}  \partial p / \partial x $ is a geostrophic velocity in the $y$-direction, $f$ is the Coriolis frequency, $g$ the gravitational acceleration, $\rho=\rho_0 + \rho(y,z)$ the density, $p$ the pressure field and $N^2 = - \frac{g}{\rho_0} \frac{\partial \rho}{\partial z}$ is the buoyancy frequency.

For a small perturbation $q'$ ($q=\overline{q}(y) + q'(x,z,t)$) on a  time-independent background PV distribution $\overline{q}$ the linearized stability problem is formulated as 
\begin{equation}
  \label{eq:listabprob}
  \frac{\D }{\D t} q' + v \frac{\partial \overline{q}}{\partial y} = 0
\end{equation}
with the perturbation velocity $v$.

To keep the formalism straightforward, only an interior solution at a distance from the boundaries in the $y$ direction is considered. The interior region is wide enough to permit the growth of waves and to justify the absence of Kelvin-wave-like boundary dynamics, but sufficiently narrow to justify the approximation of a small relative change in bottom depth: $h=h(y) \ll H$.

The mathematical problem posed by Equation (\ref{eq:listabprob}) becomes tractable if there are no gradients in the basic-state PV field. In this case $\partial \overline{q}/ \partial y = 0$, $U=U_0 z$ and the conservation law of Equation (\ref{eq:listabprob}) becomes 
\begin{equation}
  \label{eq:pvcons:zeropvbg}
  \frac{ \D }{ \D t} q' = 0,
\end{equation}
i.e., $q'$ is constant following a parcel. 

On the upper boundary, the condition for the vertical velocity is $w|_{z=H} = 0$, whereas at the lower boundary  impermeability of the boundary yields $w|_{z=h} = v \partial h / \partial y$. For small deviations from the background state and adiabatic flow, the boundary condition is obtained from the buoyancy equation $\D \rho / \D t = 0$. With the geostrophic advection operator 
and the hydrostatic approximation $\frac{\partial p}{\partial z} = - \rho g$, we arrive at the equation to be satisfied on both horizontal boundaries:
\begin{equation}
  \label{eq:buoy:def}
 \rho N^2 w + \frac{ \D }{ \D t} \frac{ \partial p }{\partial z} = 0.
\end{equation}

Equation (\ref{eq:buoy:def}) may be recast in a more illuminating form by defining the isopycnal slope as 
\begin{eqnarray}
  \label{eq:def:isopslope}
  \sigma = - (\partial \rho / \partial y ) (\partial \rho / \partial z)^{-1} = f N^{-2} \frac{\partial U}{ \partial z} = \frac{f}{N} Ri^{-\frac{1}{2}},
\end{eqnarray}
where $\partial U / \partial z$ has been evaluated geostrophically as $ g (\rho f)^{-1} \frac{\partial \rho}{\partial y}$ and $Ri$ is the geostrophic Richardson number. 

From Equation (\ref{eq:pvcons:zeropvbg}) and an application of Equation (\ref{eq:buoy:def}) on both boundaries, a linear stability problem for a perturbation ($p = p_0(y,z) + p'(x,z,t)$) is obtained:
\begin{eqnarray}
  \label{eq:buoy:pert:laplace:sigma}
  \left( \frac{\partial}{\partial t} + \frac{\sigma N^2}{f} z \frac{\partial }{\partial x} \right) \left(  \frac{ \partial^2 p' }{\partial x^2} + \frac{\partial }{\partial z} \left( \frac{f^2}{N^2} \frac{\partial p'}{\partial z} \right) \right) &=& 0 \\
  \label{eq:upbc:sigma}
  \frac{f}{N^2} \frac{\partial^2 p'}{\partial z \partial t} + \sigma  z \frac{\partial^2 p'}{\partial z \partial x} - \sigma  \frac{\partial p'}{\partial x}  &=& 0 \qquad z=H \\
  \label{eq:lowbc:sigma}
  \frac{\partial h}{\partial y} \frac{\partial p'}{\partial x} + \frac{f}{N^2} \frac{\partial^2 p'}{\partial z \partial t } + \sigma  z \frac{\partial^2 p'}{\partial x \partial z} - \sigma \frac{\partial p'}{\partial x} &=& 0. \qquad z=0 
\end{eqnarray}

The lower boundary condition should be evaluated at $z=h$, but for small Rossby numbers it can be linearized around $z=0$ \citep{Blumsack.Gierasch.1972}; otherwise the baroclinic and barotropic parts are not separable. In dimensional form, the condition is that $h$ be small compared to $H$. At this level of approximation, the bottom slope does not play a role for the horizontal momentum balance, and thus Kelvin-like boundary waves are filtered out from the problem.

Taking $N^2$ constant, the variables are separated by writing 
\begin{equation}
  \label{eq:phisepvar}
  p' = \phi(z) e^{ik(x-ct)},
\end{equation}
which results in eigenvalue equations for the equation set (\ref{eq:buoy:pert:laplace:sigma})--(\ref{eq:lowbc:sigma}):
\begin{eqnarray}
  \label{eq:eig:laplace}
  (U-c) \left( \frac{f^2}{N^2} \frac{\partial^2 \phi}{\partial z^2} - k^2 \phi \right) &=& 0 \\
  \label{eq:eig.upbc}
  \left( \sigma  H - \frac{f}{N^2} c \right) \frac{\partial \phi (H) }{\partial z} - \sigma  \phi(H) &=& 0 \\
  \label{eq:eig.lowbc}
  - \frac{f}{N^2} c \frac{\partial \phi(0) }{\partial z} + \left( \frac{\partial h}{\partial y} - \sigma  \right) \phi(0) &=& 0. 
\end{eqnarray}

We look for solutions to Equation (\ref{eq:eig:laplace}) of the form
\begin{equation}
  \label{eq:lap:sol}
  \phi = C_1 e^{- \frac{k N z}{f} } + C_2 e^{ \frac{k N z}{f} } 
\end{equation}

Non-trivial solutions for the system of equations (\ref{eq:eig.upbc})--~(\ref{eq:eig.lowbc}) must be found when the determinant of the associated coefficient matrix is zero. This condition yields a somewhat complicated characteristic polynomial, which can be simplified by defining a number of non-dimensional variables.

The Rossby height $H_R = f (kN)^{-1}$ can be used to define the height ratio $r_h \equiv H H_R^{-1} = H k N f^{-1}$, which is also the non-dimensional wave number. The inverse of the buoyancy frequency times the isopycnal slope, $ (N |\sigma|)^{-1}$, appears as a natural time scale for the problem (cf.\ \citet{Rhines.edge.1970}) and should be much larger than $f^{-1}$ for the approximations leading to Equation (\ref{eq:qgpveq}) to be valid. With this time scale, the frequency $\omega=ck$ can be non-dimensionalised as 
\begin{eqnarray}
  \label{eq:def:ndfreq}
  \tilde{\omega} = \frac{\omega }{ N \sigma  } = \frac{\omega }{ f Ri^{1/2}} = \frac{c}{H_R \partial U/\partial z }.   
\end{eqnarray}
One possible physical interpretation of $\tilde{\omega}$ is as the ratio between the phase speed and the background current velocity at one Rossby height above the bottom.

A further notational simplification is obtained by defining the ratio between the bottom and isopycnal slopes as $\Delta = \frac{\partial h }{ \partial y} \sigma^{-1}$, which leads to the following characteristic equation:
\begin{eqnarray}
  \label{eq:omega.charpoly}
   (1-\Delta) \left( (1- e^{2 r_h}) +(1+e^{2 r_h} ) r_h \right) + \nonumber \\ 
+ \left( \Delta (1 + e^{2 r_h} ) + (1-e^{2 r_h}) r_h \right) \tilde{\omega} - (1- e^{2 r_h} ) \tilde{\omega}^2 = 0
\end{eqnarray}
with the solution for $\tilde{\omega}$ is obtained as
\begin{eqnarray}
  \label{eq:omegaN:sol}
\tilde{\omega} =  \frac{1}{2(e^{2r_h}-1)} \biggl( -\Delta(e^{2 r_h} +1) + r_h(e^{2 r_h} -1) ±  \nonumber \\
 ±  \left. \sqrt{ \left( 2 - \Delta + r_h\right)^2 + e^{4 r_h} \left( -2 + \Delta + r_h \right)^2 + 
          2 e^{2 r_h} \left( -4 + \Delta \left( 4 + \Delta \right)  - {r_h}^2 \right) } \right).
\end{eqnarray}

\subsection{Solution properties}
\label{sec:solut-prop-small}

The standard Eady problem is recovered by setting $\Delta=0$. It is readily found that in this case imaginary values are possible only for $0<r_h < 2.39936$, i.e. the upper and lower boundaries must not be more than about 2.4 Rossby heights apart \citep{Eady.1949}. 

\begin{figure}[tbp]
  \begin{center}
    (a) \includegraphics[width=0.43\linewidth]{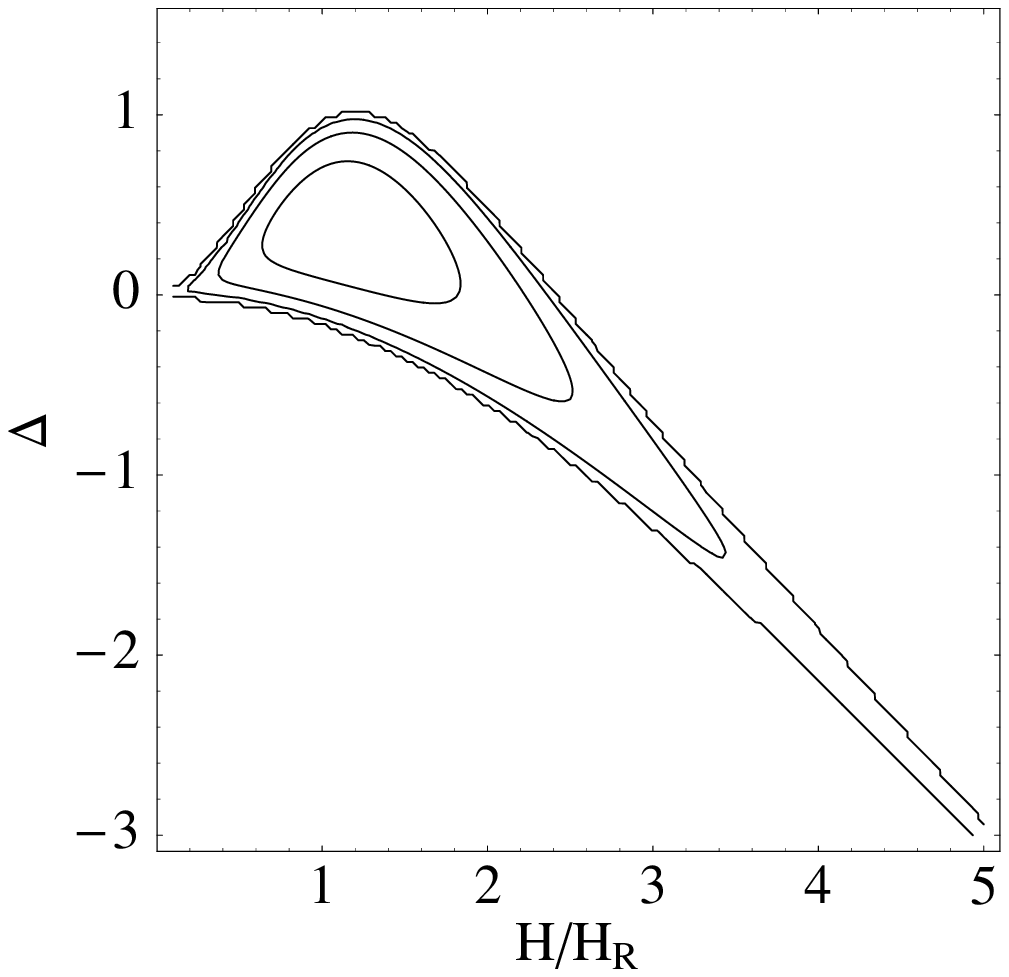}
    (b) \includegraphics[width=0.43\linewidth]{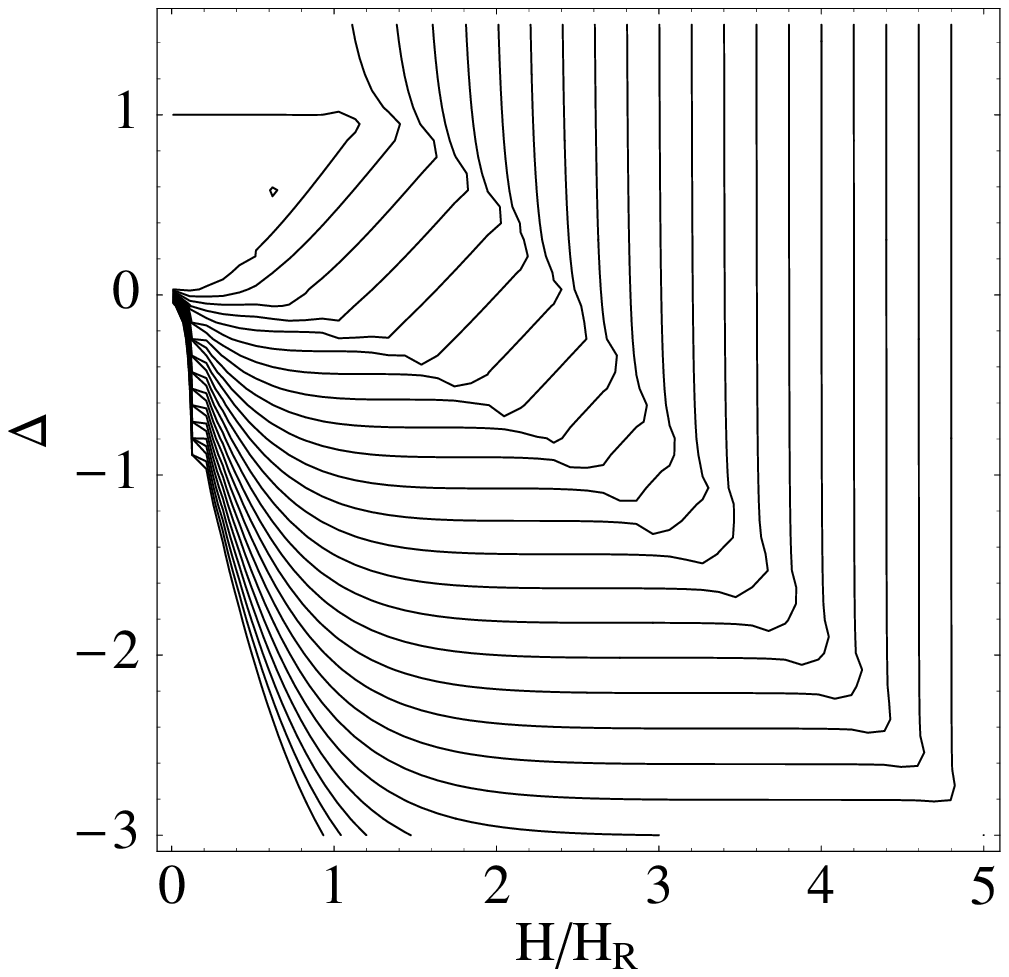}
    \caption{(a) Non-dimensional growth rate $\Im (\tilde{\omega}_{+} )= c k N ( f \partial U / \partial z )^{-1}$ as a function of the height ratio $r_h\equiv H H_R^{-1}  = H k N f^{-1}$ and the slope parameter $\Delta = \frac{\partial h }{ \partial y} \sigma^{-1}$. Contours 0, 0.1, 0.2 and 0.3 are drawn. (b) Non-dimensional frequency ($\Re (\tilde{\omega_{+}})$) as a function of $r_h$ and $\Delta$. Contours increase from zero in the upper left hand corner with 0.2 intervals.}
    \label{fig:omega:real}
    \label{fig:omega:imag}
  \end{center}
\end{figure}

With a sloping bottom boundary, the imaginary part of $\tilde{\omega}$ for a range of interesting $\Delta$-values is given in Figure \ref{fig:omega:imag}(a), where unstable solutions only are possible within the contoured area. For opposing isopycnal and bottom slopes, there is a strong stabilisation of long waves, while the short-wave cutoff moves to even shorter waves. A narrow unstable region remains, however.

For slopes in the same direction, the maximal growth rate initially increases as $\Delta$ grows, before decreasing as $\Delta \to +1$ (the maximum of $\tilde{\omega}$ is 0.366904 for $r_h = 1.13777$, $\Delta = 0.398331$, as opposed to 0.309817 in the Eady case when $r_h=1.60609$). For $\Delta>1$, i.e. when the \emph{isopycnal slope is smaller than bottom slope}, no unstable solutions are to be found, a result that is formally demonstrated in Section \ref{sec:stabcond}.

The real part of $\tilde{\omega}$ in Figure \ref{fig:omega:real}(b)  also shows intriguing characteristics and seems to have escaped discussion. (Figure \ref{fig:omega:imag}(a) is  essentially similar to Figure 2 of \citet{Blumsack.Gierasch.1972} but these authors appear to have neglected the contribution from the discriminant to the real part of the solution. Note that \citet{deSzoeke.1975} gives a slightly different stability diagram, apparently because of his bounded $y$-domain.) It is readily seen that in the unstable wedge, the waves are only weakly dispersive since $\tilde{\omega}$ increases in an approximately linear fashion towards $r_h \propto  k$, i.e. $c \approx {\rm const}$. However, outside the unstable wedge and disregarding the very longest waves, the waves are dispersive, implying that wave packages with different wave numbers would propagate away from each other. 

\begin{figure}[tp]
  \begin{center}
   (a) \includegraphics[width=0.7\linewidth]{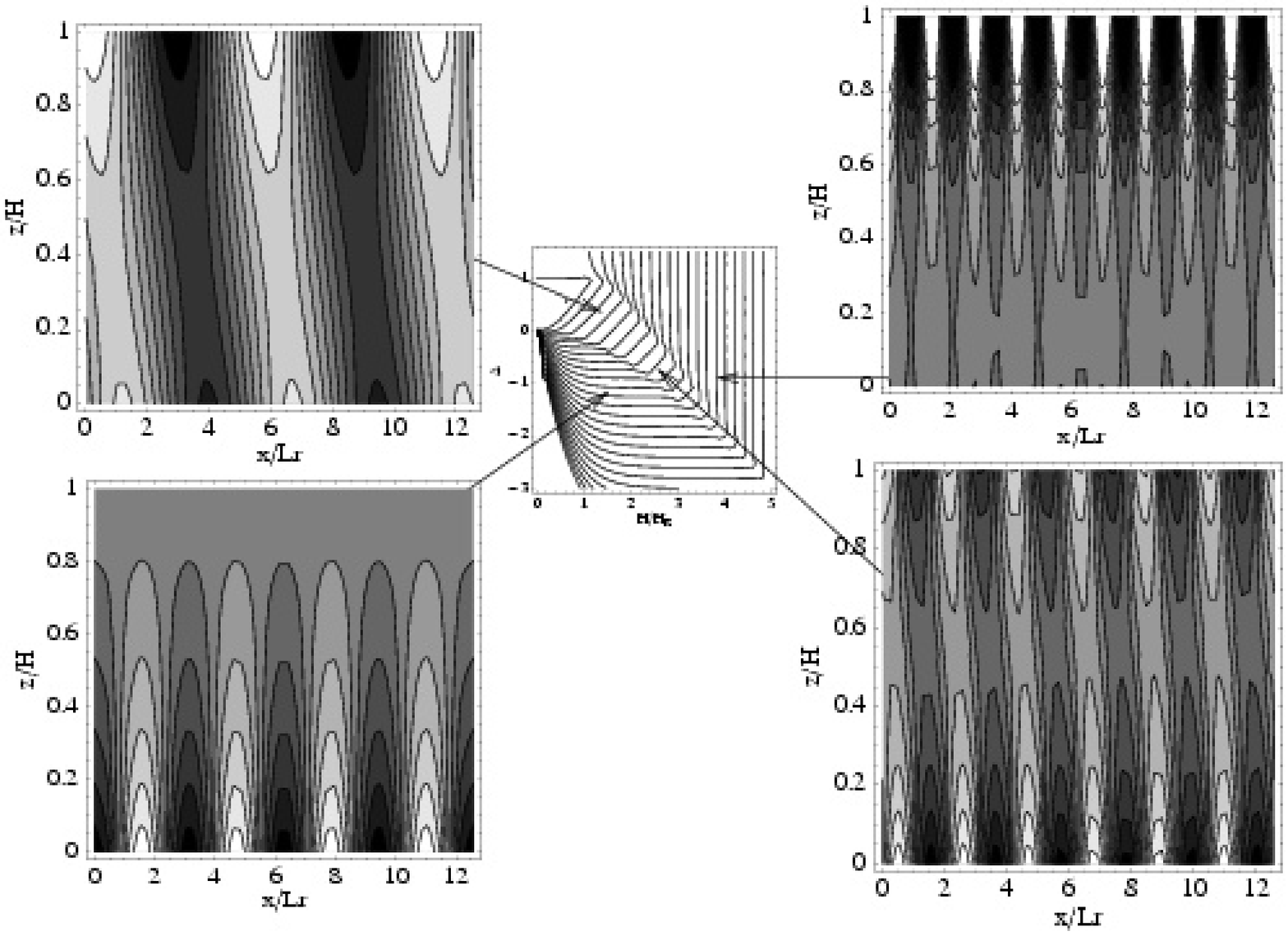} \\
   (b) \includegraphics[width=0.7\linewidth]{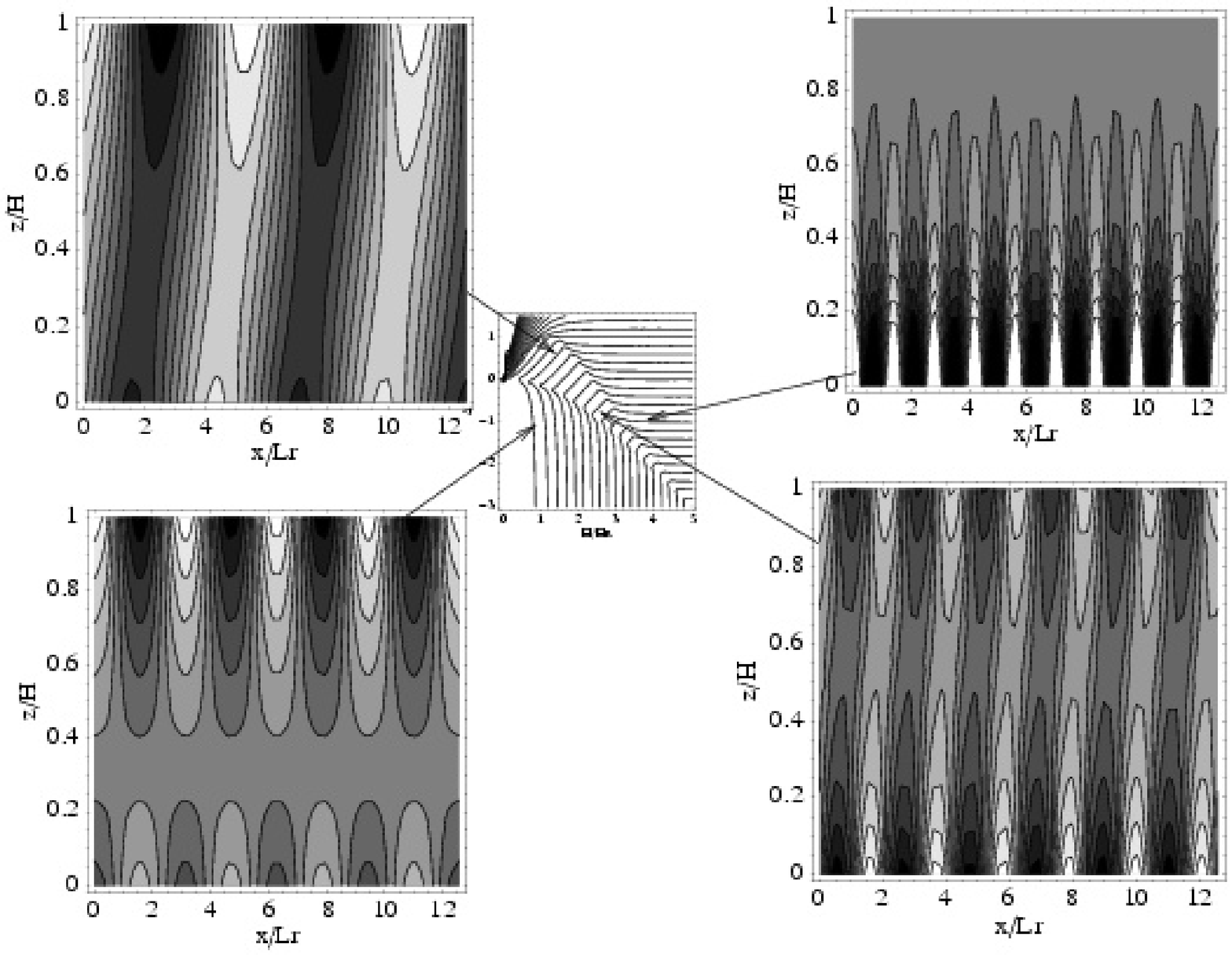}
    \caption{\label{fig:regimes}The structure within a constant, i.e., $p' C_1^{-1}$, of growing (a) and decaying (b) perturbations in the ($r_h$, $\Delta$) --space. The vertical axis is scaled with the domain height, $\tilde{z}=z H^{-1}$, and the horizontal is scaled with the Rossby radius, $\tilde{x}=x L_{R}^{-1} = x f (N H)^{-1}$ with a total extent of $4\pi$. Outside the region of instability, the waves are confined to either boundary. Mean shear is to the right (positive $x$ direction) in the figure}
  \end{center}
\end{figure}

The different propagation characteristics are more understandable in the light of the structure of the disturbances (Figure \ref{fig:regimes}). It is recognised that the dispersion properties are determined by the boundary at which the disturbance is trapped; on the lower boundary, $\tilde{\omega} \propto \Delta $, whereas on the upper boundary $\tilde{\omega} \propto r_h $. 

The wave properties and the resulting instability may be interpreted as the interaction between two boundary waves (e.g. \citet{Bretherton.QJRMC.1966b}). For $\tilde{\omega}_{+}$ and opposing isopycnal and topographic slopes ($\Delta < 0$), the long waves are confined to the lower boundary. For increasing wavenumbers, the upper boundary comes into play, and the perturbations on both boundaries start to interact and grow in time, because the shear maintains their phase difference while the mutually induced velocity field tends to diminish it. For even larger wavenumbers, the contribution from the lower boundary vanishes and what remains is a stable wave on the upper boundary. 

The most unstable solution differs somewhat in character from the Eady wave, since it is more confined to the upper boundary. This is caused by the sloping bottom boundary amplifying the vorticity contributions for small perturbations. The result is that in the interaction regime, larger perturbations on the upper boundary are permitted, resulting in an overall amplification of the instability.

For $\tilde{\omega}_{-}$, the situation is reversed: long waves are found on the upper boundary, short ones on the lower boundary, and in the interaction zone the upper boundary perturbation is ``upstream'' of the lower boundary, which gives rise to a destructive interaction.

Moreover, there is a narrow regime of negative frequencies for $0.4<\Delta< 0.6$ and $0.4<r_h<0.6$. The possible relevance of this regime to a more realistic problem has not been discussed, but it appears as though the edge-wave phase speed exceeds the Eady-wave phase speed in this regime, resulting in wave progression  with the coast on the right.

\section{Eady problem with Ekman pumping on the lower boundary}
\label{sec:eady-problem-with-Ekman}

The effects of a viscous boundary layer on the interior balance may be represented by diagnosing the vertical velocity at the top of the bottom boundary layer using the steady-state Ekman transport, 
\begin{eqnarray}
  \label{eq:friction:ekman:We}
  w_E = \left( \frac{\nu }{2 f } \right)^{1/2} \left( \frac{\partial v}{\partial x} - \frac{\partial u_g}{\partial y} \right) = \frac{1}{\rho f } \left( \frac{\nu }{2 f } \right)^{1/2} \left( \frac{\partial^2 p'}{\partial x^2} + \frac{\partial^2 p'}{\partial y^2} \right),
\end{eqnarray}
which in the absence of $y$-variations can be written as $ w_E =\frac{1}{\rho f } \left( \frac{\nu }{2 f } \right)^{1/2} \frac{\partial^2 p'}{\partial x^2} $. Hence, the lower boundary condition of Equation (\ref{eq:lowbc:sigma}) can be reformulated as
\begin{eqnarray}
  \label{eq:friction:ekman:lowbc}
  \left( \frac{\nu }{2 f } \right)^{1/2}  \frac{\partial^2 p'}{\partial x^2} + \frac{f}{N^2} \frac{\partial^2 p'}{\partial z \partial t } + \sigma  z \frac{\partial^2 p'}{\partial x \partial z} - \sigma  \frac{\partial p'}{\partial x} &=& 0. \qquad z=0   
\end{eqnarray}

The solution procedure of Section \ref{sec:fullsol} can now be repeated, resulting in a partly imaginary characteristic polynomial
\begin{eqnarray}
  \label{eq:charpoly:Ekman}
  e^{2\,{r_h}}\,\left( 2 - \mi\, \Delta_\nu  - 
      2\,{\tilde{\omega}} \right) \,\left( -1 + {r_h} - {\tilde{\omega}} \right)  + 
   \left( 1 + {r_h} - {\tilde{\omega}} \right) \,
    \left( 2 - \mi\, \Delta_\nu  + 2\,{\tilde{\omega}} \right) = 0,
\end{eqnarray} 
where $\Delta_\nu =  k \sigma^{-1}  \sqrt{ \nu  f^{-1} } \equiv \hat{E}_\nu \sigma^{-1} $. $\hat{E}_\nu = k \sqrt{\nu  f^{-1} }$ is an Ekman number based on the horizontal wave number rather than the depth scale; note that $\hat{E}_\nu$ is also a slope-like parameter in that it is the ratio of a vertical to a horizontal length. $\Delta_\nu$ is related to the ratio between the Ekman and Rossby-Kibel numbers \citep[cf. their definition of $Q$]{Williams.Robinson.1974.geneady}.

The solution of Equation (\ref{eq:charpoly:Ekman}) is
\begin{eqnarray}
  \label{eq:Ekmansol}
\tilde{\omega} = \left( 4 \left( e^{2 r_h} -1 \right) \right)^{-1}
\Biggl( \mi \Delta_\nu \left( e^{2 r_h} +1 \right)  - 
      2 r_h \left( e^{2 r_h} -1 \right) ± \nonumber \\ 
     \pm \biggl( - {\Delta_\nu }^2 { \left( e^{2 r_h} +1 \right) }^2   + 4 \mi \Delta_\nu 
           \left( e^{2 r_h } -1 \right) 
           \left( 2 + e^{2 r_h } \left( r_h -2 \right)  + r_h \right)  + \nonumber \\
          + 4 \left( e^{2 r_h } -1 \right) 
           \left( e^{2 r_h } {\left( r_h -2 \right) }^2 - {\left(  r_h + 2 \right) }^2 \right) \biggr)^{1/2} \Biggr) 
\end{eqnarray}

\begin{figure}[tp]
  \begin{center}
   (a) \includegraphics[width=0.43\textwidth]{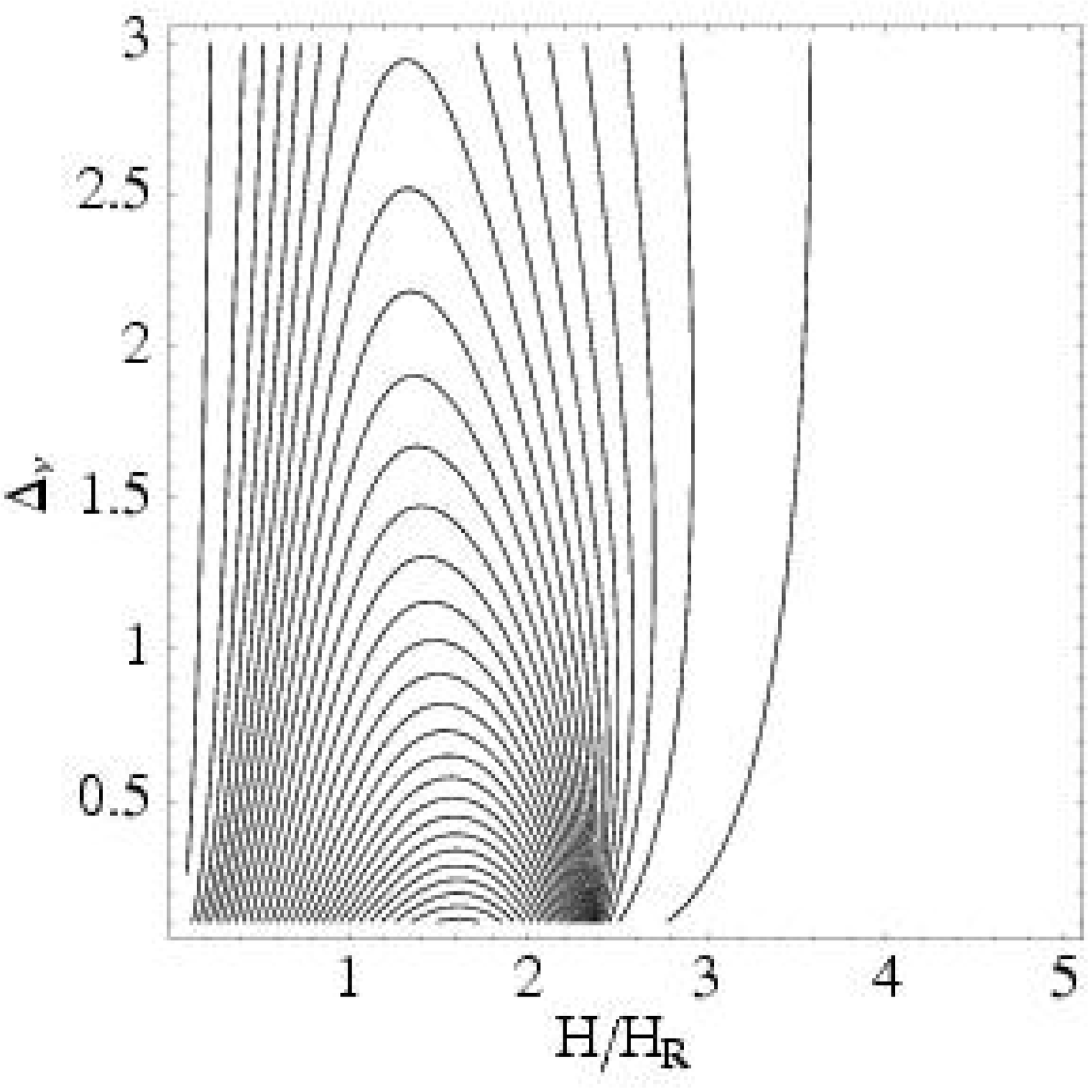}
   (b) \includegraphics[width=0.43\textwidth]{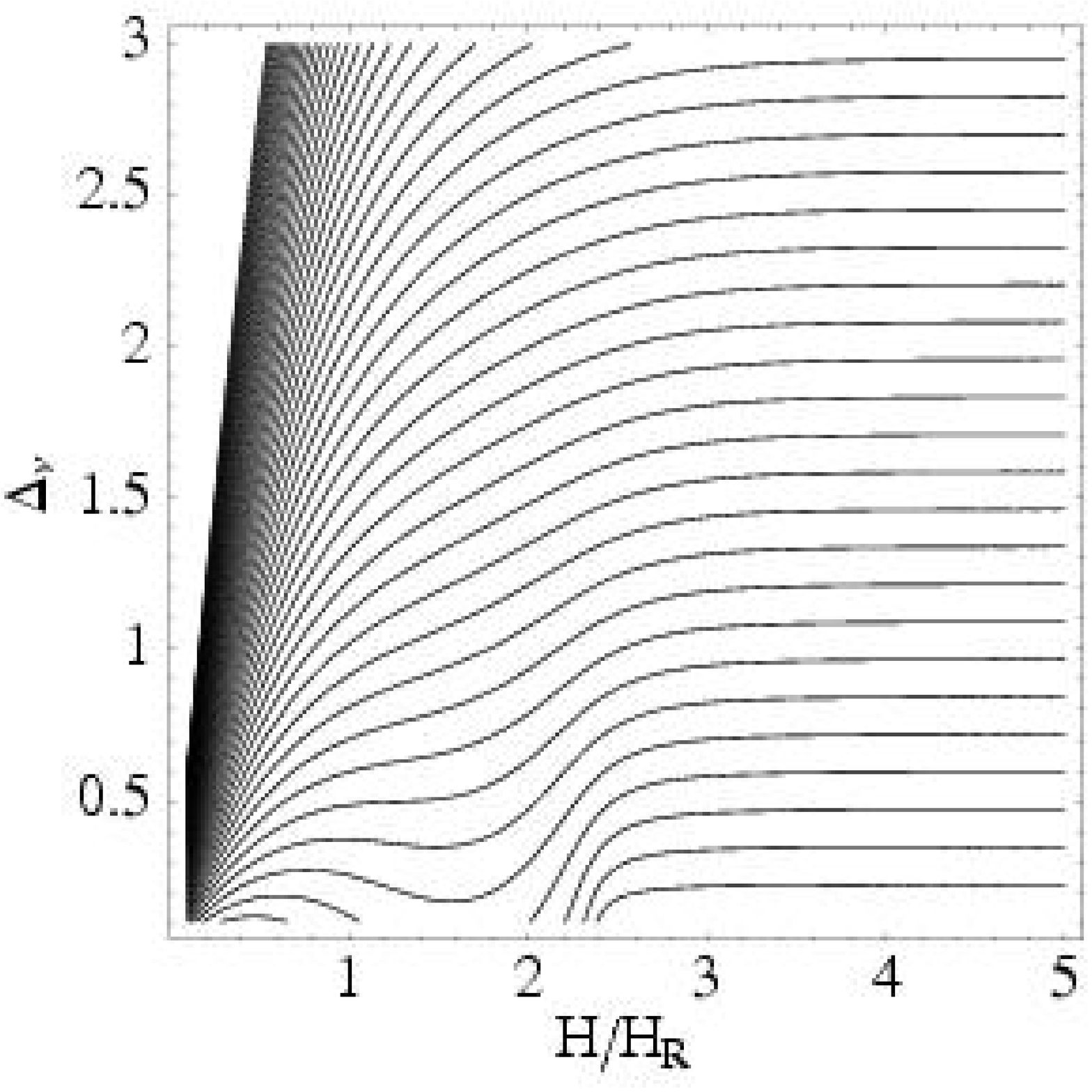}
    \caption{Growth rate for (a) the growing solution $\Im [ \tilde{\omega}_{+} ]$ and (b) the decaying solution $\Im [ \tilde{\omega}_{-} ]$ to the Eady-Ekman problem,  as a function of the non-dimensional friction $\Delta_\nu$ and boundary separation $r_h = H H_R^{-1}$. The decaying solution is obtained with negative $\Delta_\nu $ due to the symmetry of square roots of complex numbers.}
    \label{fig:stab:ekaman}
  \end{center}
\end{figure}

The associated stability properties are shown in Figure \ref{fig:stab:ekaman}. Note that since a change in the sign of the isopycnal slope affects both $\Delta$ and the \emph{dimensional} version of $\omega$ (Eq.~(\ref{eq:def:ndfreq})), the $\Delta_\nu < 0$ part of the $\tilde{\omega}_{+}$  solution is actually decaying, and vice versa for  $\tilde{\omega}_{-}$; the  positive and negative solutions switch roles when $\Delta_\nu $ changes sign and hence only the $\Delta_\nu \geq 0$ region is shown.

\begin{figure}[tp]
  \begin{center}
   (a) \includegraphics[width=0.8\textwidth]{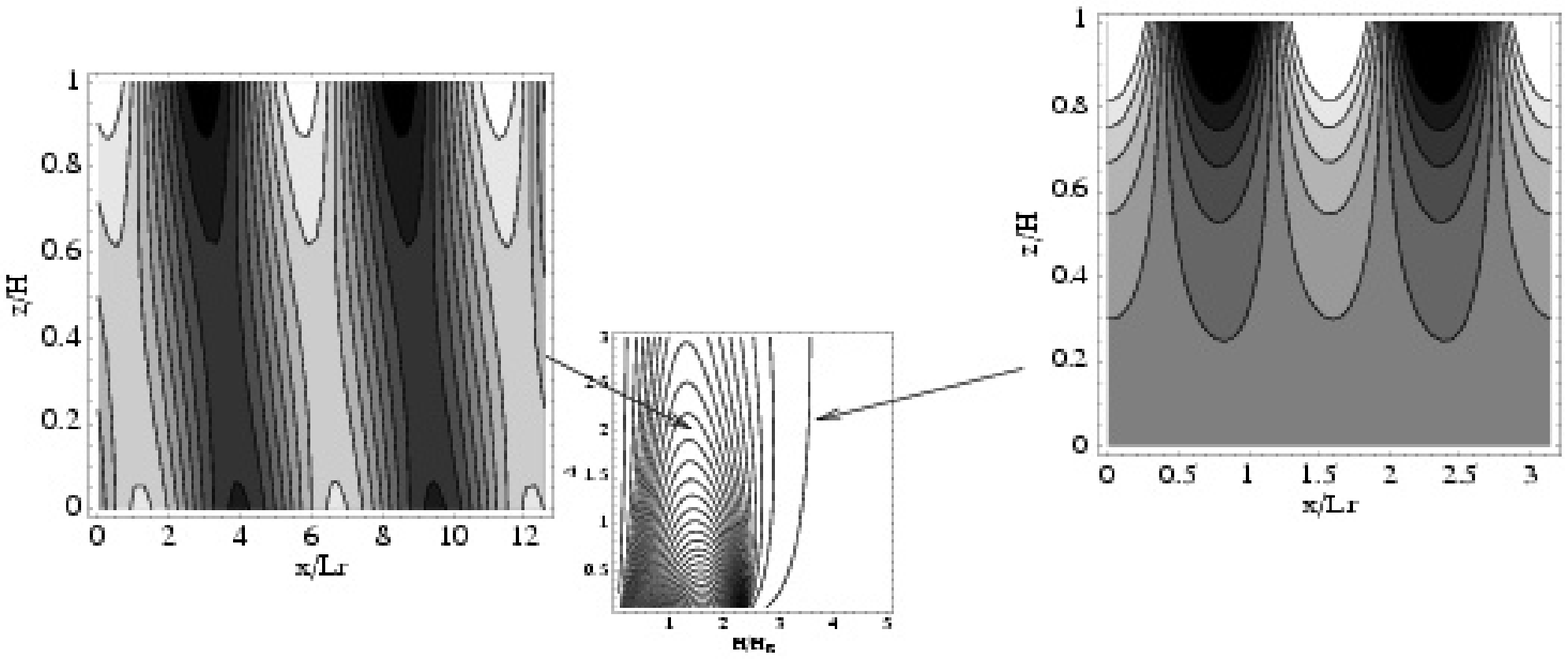} \\
   (b) \includegraphics[width=0.8\textwidth]{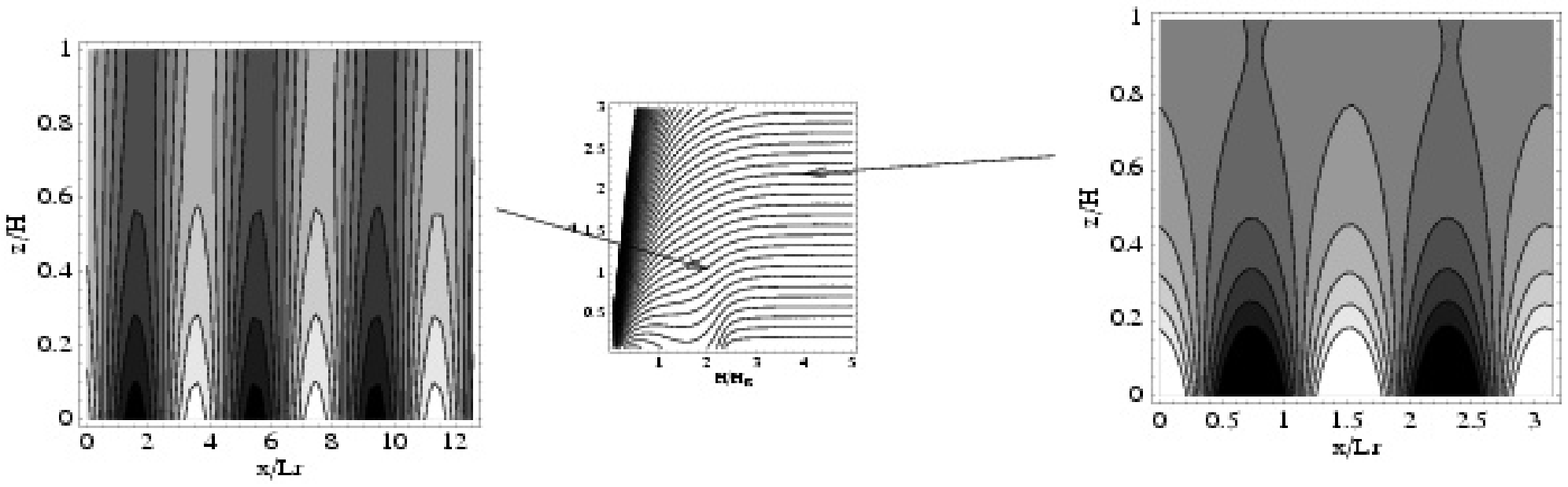}
    \caption{Structure of (a) positive ($\tilde{\omega}_{+}$) ( ($r_h=1.6, \Delta_\nu=5$), ($4.0,2.0$)),  and (b) negative ($\tilde{\omega}_{-}$) solutions (($1.6 ,1$) and ($4.0, 2.0$)) in the Eady-Ekman problem. Note the change in $x$-axis length for rightmost figures.}
    \label{fig:struct:growing}
  \end{center}
\end{figure}

The stability characteristics of $\tilde{\omega}_{+}$ in Figure \ref{fig:stab:ekaman}(a) can again be understood from the structure of the perturbations, as presented in Figure \ref{fig:struct:growing}(a). For the short waves that are stable in the Eady problem ($r_h > 2.4$), there is a  change in the character of the solution with  changes in $\Delta_\nu$. The waning Eady solution, which is bottom-trapped in the short-wave range, is barotropised and becomes strongly decaying by increased Ekman pumping. The damping increases exponentially in the long-wave limit $r_h \to 0+$.

The growing solution is top-trapped and destabilized for short wavelengths due to baroclinisation. Moreover, in the range $r_h > 2.4$, the maximum growth rate is always found for non-zero values of  $\Delta_\nu$, since the Eady solution $\Delta_\nu=0$ is stable.

\begin{figure}[tp]
  \begin{center}
    (a) \includegraphics[width=0.43\textwidth]{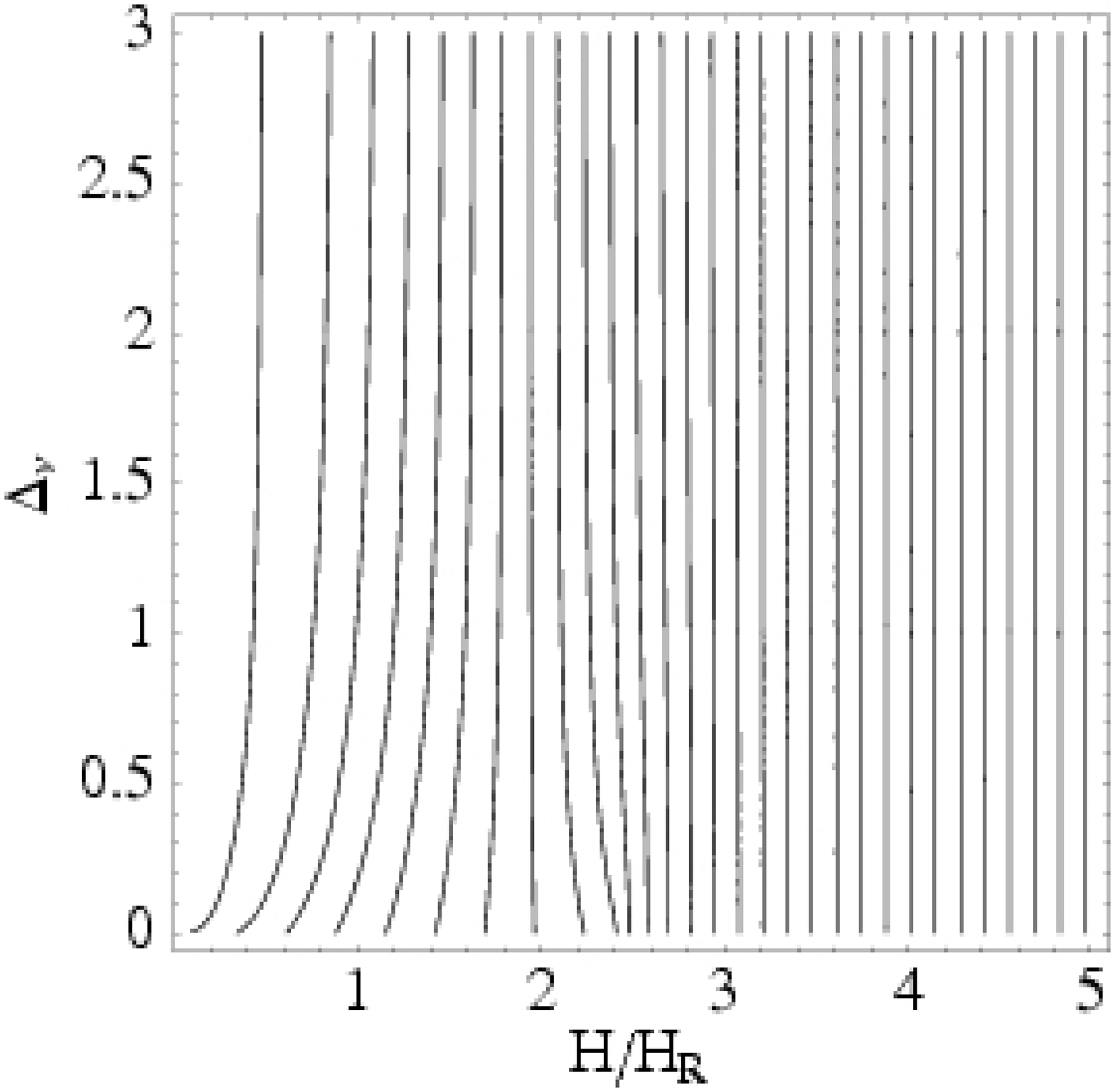}
    (b) \includegraphics[width=0.43\textwidth]{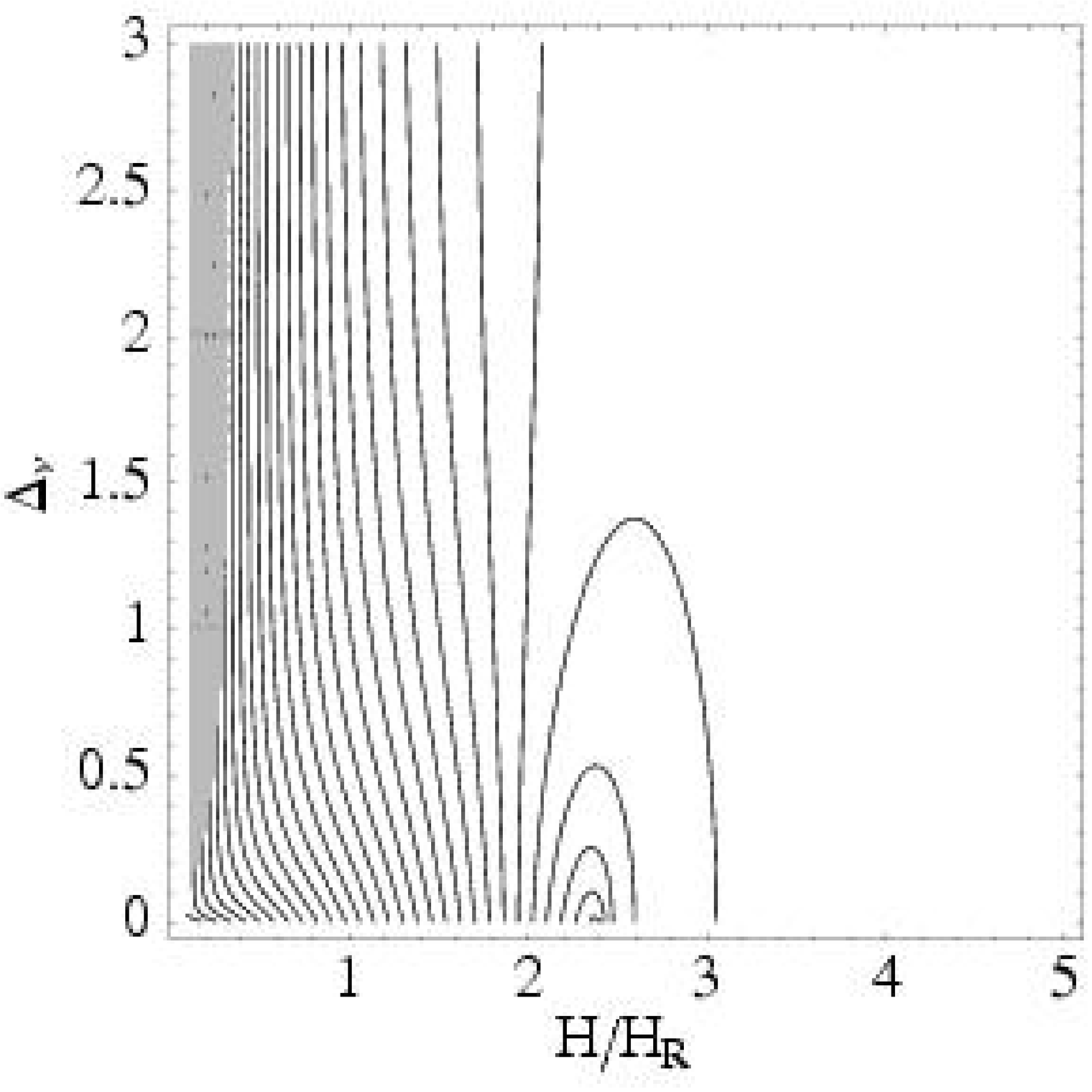}
    \caption{Real part of $\tilde{\omega}$ for (a) the growing and (b) the decaying disturbance in the Eady-Ekman problem.}
    \label{fig:cphase}
  \end{center}
\end{figure}

The stability properties also manifest themselves in the phase speed of the waves as an increase (decrease) for decaying (growing) long waves, cf.~Figure~\ref{fig:cphase}.

\section{Discussion}
\label{sec:discuss}
\label{sec:discuss:gener}

In the previous sections the effect of the lower boundary condition on the phase speed of baroclinic disturbances and their vertical trapping was found to be substantial. Keeping in mind the basic mechanism of instability in the Eady problem, we propose that any non-standard boundary condition that modifies the boundary-wave phase speed will affect the stability properties in a rather fundamental way. Some possible consequences of varying boundary conditions and their interpretations are discussed in the following subsections.

\subsection{The edge wave---Eady wave duality}
\label{sec:discuss:phasespeed}

The phase speeds of the solutions to the bottom-slope problem when the disturbances are trapped to either boundary may be approximated from Equation (\ref{eq:omegaN:sol}) as
\begin{eqnarray}
  \label{eq:phasespeed:approx:ompos}
  \lim_{r_h \to\infty} \tilde{\omega}_{+} = r_h -1 \\
  \lim_{r_h \to\infty} \tilde{\omega}_{-} = 1-\Delta, 
\end{eqnarray}
which yields the following dimensional results:
\begin{eqnarray}
  \label{eq:phasespeed:top:lower}
  c_{upper} =  \left(U(H) - \frac{\partial U}{\partial z} H_R  \right) = \sigma( \frac{N^2 H}{f} - \frac{N }{k} ), \\
  c_{lower} =  \left(\frac{\partial U }{\partial z} - \frac{N^2}{f} \frac{\partial h}{\partial y}  \right) H_R = (\sigma - \frac{\partial h }{\partial y} ) \frac{N}{k}.
\end{eqnarray}
This result for non-interacting boundary waves could also have been obtained by inserting the exponential solution of Equation (\ref{eq:lap:sol}) separately into Equation (\ref{eq:eig.lowbc}) or Equation (\ref{eq:eig.upbc}) and solving for $c$, yielding a wave solution for one boundary only.
 
For the topographically modified Eady problem with positive shear, the lower boundary phase speed is reduced by a boundary with a positive slope, because the slope generates a quasigeostrophic edge wave that tends to travel in the Kelvin-wave direction, opposite to the Eady wave and therefore decelerating it in the case of positive shear, but accelerating it if $\Delta<0$. Indeed, even negative phase speeds are possible on the lower boundary.

\begin{figure}[tp]
  \begin{center}
    (a) \includegraphics[width=0.43\textwidth]{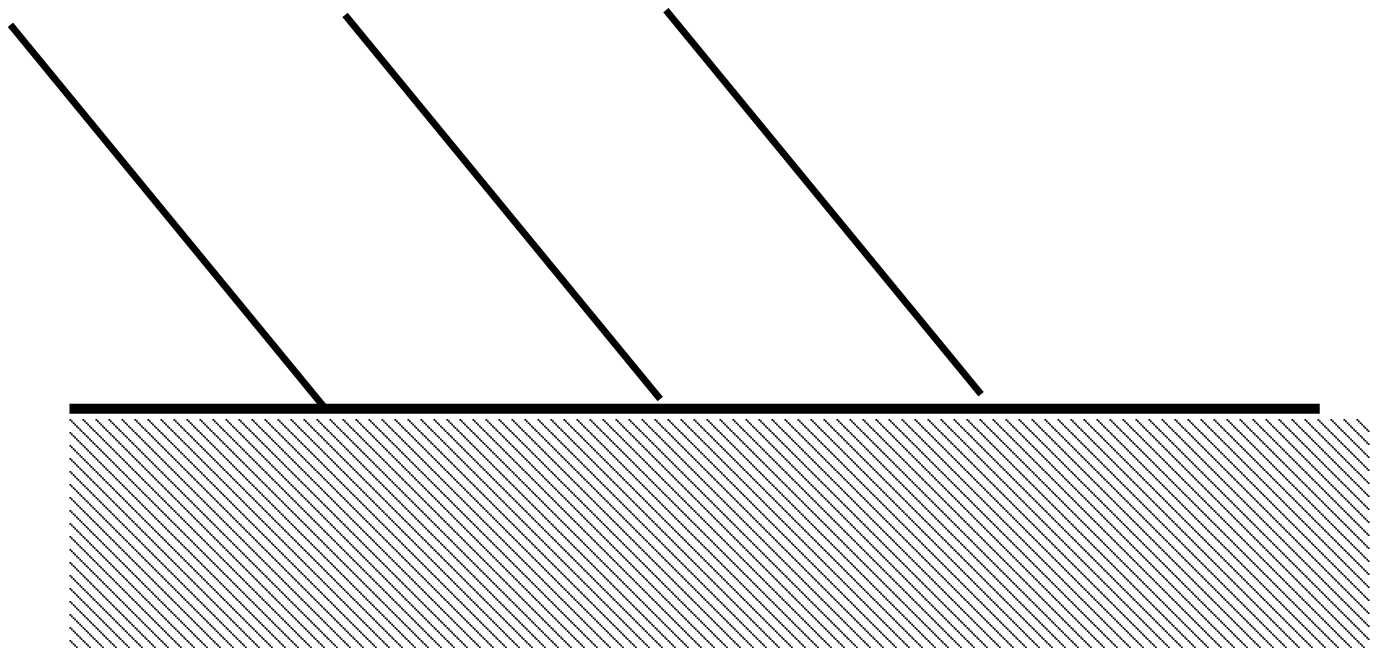} 
    (b) \includegraphics[width=0.43\textwidth]{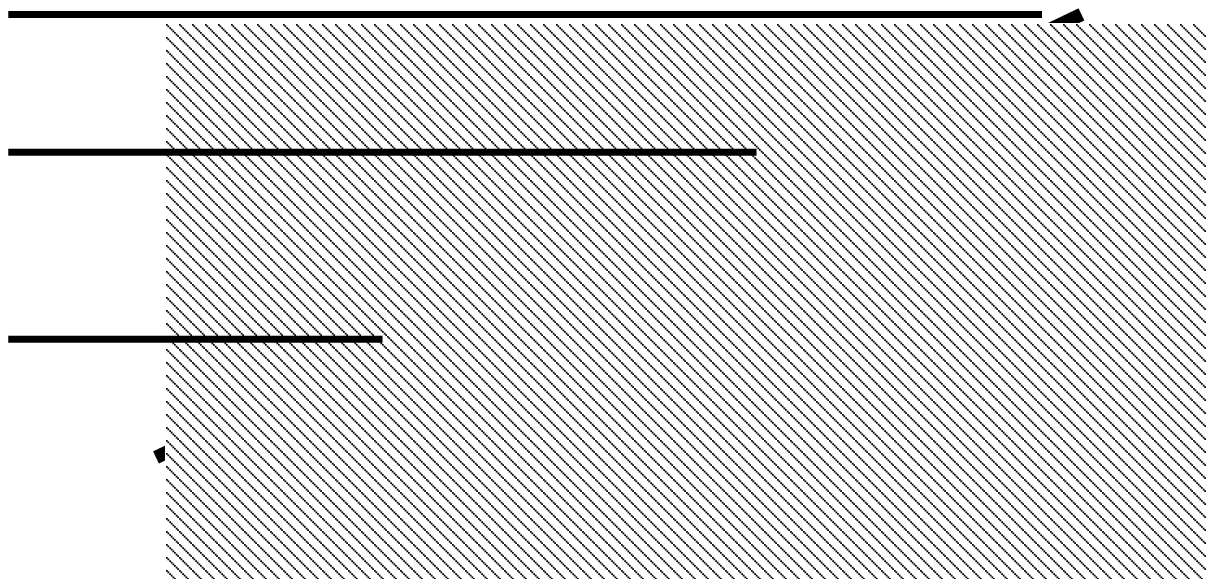} 
    \caption{Duality between stable Eady waves with a flat bottom but sloping isopycnals (a) and edge waves with flat isopycnals but sloping bottom (b): the waves on the boundary have the same phase speed, only the angle between the boundary and the isopycnals matters}
    \label{fig:dual:eadyedge:ill}
  \end{center}
\end{figure}

Within this approximation, there is a duality between Eady waves with sloping isopycnals on a flat bottom and edge waves  with horizontal isopycnals on a sloping bottom (cf.\ \citet{Rhines.edge.1970}). The same phase speed as with $\partial h / \partial y = 0$, $\sigma = \sigma_0$ for $c_{lower}$ is obtained with $\sigma =0 $ if $\partial h / \partial y = - \sigma_0 $. The duality is illustrated in Figure \ref{fig:dual:eadyedge:ill}. For $\sigma=0$, there are no unstable solutions.

Note also, that the phase speed on the lower boundary in the absence of an isopycnal slope, $c_{lower} = - \frac{\partial h }{\partial y}  \frac{N}{k}$, is frequently referred in the literature as the Nof speed, being the speed at which eddies progress on a sloping boundary. An explicit connection to the edge wave phase speed of \cite{Rhines.edge.1970} is appropriate in this context.

\subsection{Surface temperature gradient vs. isopycnal slope; the necessary condition for instability}
\label{sec:stabcond}

It is customary to decompose the quasigeostrophic dynamics into contributions from the interior PV and from the boundary temperature gradient (e.g.\ \citet{Hoskins.etal.1985}). \citet{Holopainen.Kaurola.1991} have  criticized this decomposition by arguing that the quantities  depend on each other, and when evaluated from observational data there is a high degree of cancellation in the inverted fields over large scales (see also \citet{Davis.comment.1993,Holopainen.Kaurola.1993}). We briefly revisit this issue from a stability perspective.

The vertical divergence $\partial / \partial z$ of the boundary condition of Equation (\ref{eq:buoy:def}) may be reformulated as
\begin{equation}
  \label{eq:buoy:def:mod}
 \left( \frac{\partial }{\partial t} + U(z) \frac{\partial }{\partial x} \right) \frac{\partial }{\partial z} \frac{ f^2 \partial p }{N^2 \partial z} = - \frac{\partial }{\partial z} ( \rho f^2 w ) - v \frac{\partial }{\partial y} \frac{\partial }{\partial z} \frac{f^2 \partial p }{N^2 \partial z} ,
\end{equation}
which, with $w= v \partial h / \partial y $, yields  
\begin{eqnarray}
 \label{eq:qgpveq:bgpv}
 \frac{ \D  }{ \D t} \frac{\partial }{\partial z}  \left( \frac{f^2 \partial p }{N^2 \partial z} + f^2 \rho h \right) =  0  \qquad z=0,H.
\end{eqnarray}

The range of validity of the Poisson equation (\ref{eq:buoy:pert:laplace:sigma}) may now be extended to the full vertical domain, if the boundary conditions are generalised to a flux, which is constant in the interior and diverges on the boundaries, so that the $y$-term of the material derivative becomes $v \frac{\partial }{\partial y}  \frac{\partial }{\partial z}  \mathcal{H}(z) \left( \frac{f^2 \partial p }{N^2 \partial z} + f^2 \rho h \right) = v \frac{\partial }{\partial y}  \delta(z)  \left( \frac{f^2 \partial p }{N^2 \partial z} + f^2 \rho h \right)$. Here $\mathcal{H}(z)$ is the Heaviside step function and its vertical derivative, the $\delta$-function, ensures that the boundary condition only appears exactly at the boundary. Thus the boundary effects may be incorporated in Equation (\ref{eq:qgpveq}) as 
\begin{eqnarray}
  \label{eq:qgpveq:deltafs}
  \frac{\D }{\D t} \tilde{q} \equiv 
\frac{\D }{\D t} \left( \frac{\partial^2 p}{\partial x^2} + \frac{\partial }{\partial z} \left( \frac{f^2 }{N^2} \frac{\partial p}{\partial z} \right)  \right) + \nonumber \\  + \frac{\D }{\D t} \left(  \delta(z) \left( \frac{f^2 \partial p }{N^2 \partial z} + f^2 \rho h \right) - \delta(z-H)   \frac{f^2 \partial p }{N^2 \partial z} \right) =0.
\end{eqnarray}
\citet{Bretherton.QJRMC.1966a} was the first author to suggest a form similar to Equation (\ref{eq:qgpveq:deltafs}), but without considering it as the divergence of a flux.

The form of PV in Equation (\ref{eq:qgpveq:deltafs}) is convenient when applying integral theorems, such as the stability theorem of \citet{Charney.Stern.1962}. This theorem requires a change in sign of $\partial \tilde{q} / \partial y$ for instability. Since there are no interior PV gradients in the Eady problem, the $y$-gradients in the boundary PV contributions can be recognized as the source of instability \citep{Bretherton.QJRMC.1966a}; compare Equation (\ref{eq:qgpveq:deltafs}) with Equation (\ref{eq:listabprob}).

It is now possible to pose the question whether the source of instability at the lower boundary can be eliminated. This proves to be the case when both contributions at $z=0$ add to zero in Equation (\ref{eq:qgpveq:deltafs}): $\frac{f^2}{N^2} \frac{\partial^2 p'}{\partial z \partial y} = - f^2 \rho \frac{\partial h}{ \partial y }$, i.e. when  
\begin{eqnarray}
  \label{eq:slopes}
  - \frac{\partial z}{\partial \rho} \frac{\partial \rho}{\partial y} =  \frac{\partial h}{ \partial y }.
\end{eqnarray}

This shows that the removal of the necessary condition for instability is possible when the \emph{the bottom slope is equal or larger than the isopycnal slope}. This possibility for stabilisation is also remarked on by \citet{Charney.Flierl.1981.analogues}. The appropriate physical explanation, however, is found by recognizing that the boundary density gradient, which is often referred to as the source of instability, is a side-effect of the isopycnal surfaces intersecting the horizontal boundary.

In the stabilized case, the isopycnals intersect only the top boundary; they do not feel the vorticity effect of the bottom because they are ``parallel'' to it. Hence, by removing the lower boundary PV source, the necessary condition for instability is also removed, and the flow is unconditionally stable, as seen in the growth rates in Figure \ref{fig:omega:real}. (\citet{Mechoso.1980} explained this removal of instability using heat-transport arguments, but since the quasi-geostrophic dynamics are determined by vertical velocity divergence, the heat fluxes are a consequence rather than a cause of the stabilisation.)

This result, strictly valid under the narrow limitations of quasi-geostrophy, makes a lot of physical sense and is therefore expected to be relevant also for non-quasi-geostrophic problems. Indeed, a parameter similar to $\Delta$ appears independently in several contexts in the literature, e.g. \citet{Swaters.jfm.1991} found a similar condition for a non-quasi-geostrophic frontal model. However, other types of instabilities that may emerge under different approximations cannot be ruled out.

For opposing isopycnal and bottom slopes, the necessary condition for instability is fulfilled, but nevertheless considerable stabilisation of in particular long waves takes place already for very small bottom slopes, cf. Figure \ref{fig:omega:real}. 

\begin{figure}[tp]
  \begin{center}
    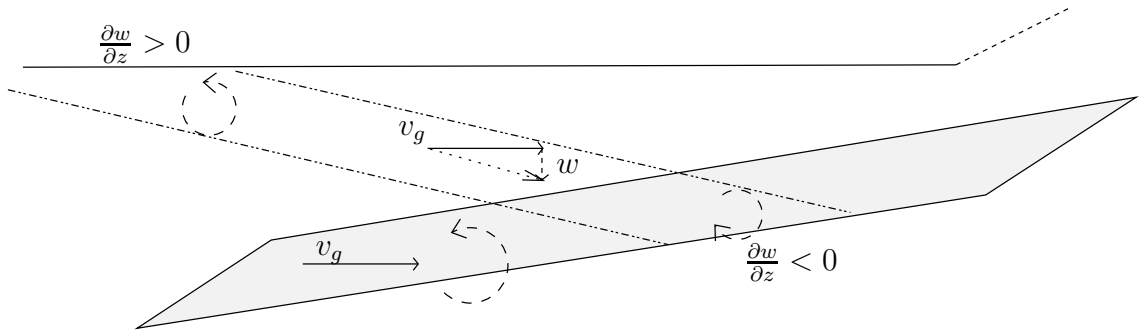
    \caption{Illustration of the divergence of vertical velocity in the Eady problem with a sloping bottom. A horizontal flow against isopycnals is converted into vertical velocity, which diverges where isopycnals intersect the boundaries, generating vorticity by vortex stretching. The bottom slope creates a divergence of its own.}
    \label{fig:ill:vertveldiv}
  \end{center}
\end{figure}

It is useful to note that the source of vorticity on the boundaries is due to their impermeability, implying $w=0$. The divergence of vertical velocity on the boundaries is the source of the $\delta$-function vorticity sheet, which makes the angle between the isopycnals and the bottom the relevant parameter for the phase propagation. The validity of this argument is born out by examining  Equation (\ref{eq:buoy:def}), which may be reformulated as
\begin{eqnarray}
  \label{eq:buoy:w:rearr}
  \rho f^2 \frac{\partial w}{\partial z} = - \frac{\D}{\D t} \left( \frac{\partial }{\partial z} \frac{f^2}{N^2} \frac{\partial p}{\partial z} \right).
\end{eqnarray}
For the advecting $y$-part of $\D / \D t$, Equation (\ref{eq:buoy:def}) yields $w =  v \sigma $, i.e. the flow of $v$ against the isopycnals causes a vertical velocity that diverges only when $\sigma$ changes or a boundary is encountered, as illustrated in Figure \ref{fig:ill:vertveldiv}. This divergence introduces a change of the interior PV, as indicated by (\ref{eq:buoy:w:rearr}). For the linearized problem, the isopycnal surfaces of the basic state are impermeable. This isentropic upgliding \citep{Hoskins.etal.1985} is exactly canceled at the boundaries in the Eady problem by the boundary-induced velocity (``boundary temperature advection'') to make the net $w$ zero. 

From a physical standpoint, it thus makes a lot of sense to divide the quasi-geostrophic flow into contributions from the horizontal vorticity ($\nabla_H^2 p$), and the modifications caused by divergence of vertical velocity ($\frac{\partial }{\partial z} \frac{f^2}{N^2} \frac{\partial p}{\partial z} + \delta$-functions), giving rise to the familiar phenomenon of vortex stretching. The natural corollary of this argument is the second flow partition suggested by \citet{Holopainen.Kaurola.1991}.

We further remark that it is customary to draw parallels between the stabilisation caused by a bottom slope and the $\beta$ effect \citep{Jungclaus.meddy.1999,Spall.Chapman.1998}. There is, however, a clear difference between the planetary vorticity gradient and the effect of topography. The bottom slope appears in the purely baroclinic problem as a $\delta$-function vorticity contribution at the boundary, whereas the $\beta$-effect is homogeneous over the whole depth. Therefore, the effect of the slope on the \emph{baroclinic} instability is fundamentally different from the $\beta$-effect, whereas for the \emph{barotropic} instability the effect is the same, but only within the linearisation permitted by $h \ll H$. Furthermore, as \citet{Charney.Flierl.1981.analogues} point out, the variation of $N$ with depth in the ocean also makes the vorticity dynamics (and hence, the stability properties) deviate from those of the simple model based on constant stratification.

\subsection{The sensitivity of eddy parameterisations}
\label{sec:sub:sensinsitivity}

The recent decade has seen a rise of interest in parameterising the effect of eddies in coarse (viz.~non-eddy-resolving) ocean models. Already a rather coarse AGCM is capable of resolving the atmospheric eddies, whereas an eddy-resolving OGCM is a major computational challenge that will be unaffordable for routine research for many years to come. 

Building on the atmospheric studies of \cite{Green.1970} and \cite{Stone.1972}, \citet{Visbeck.etal.1997} based their eddy parameterisation on the baroclinic instability growth rate in the problem of \citet{Eady.1949}. Subsequently, many studies have found that a parameterization of eddy fluxes based on the Eady growth rate can be brought to reasonable harmony with eddy-resolving 3-dimensional model results far from the lateral boundaries \citep{Visbeck.etal.1997,Jones.Marshall.1997,Spall.Chapman.1998}, though only in an ensemble-mean sense and with significant error bars \citep{Gille.Davis.1998}.

In light of the results from Sections \ref{sec:fullsol} and \ref{sec:eady-problem-with-Ekman} we may now inquire about the sensitivity of the parameterisation schemes referred to above on the details of the lower boundary, with emphasis on variations typically found in coastal-flow situations. Based on dimensional arguments (cf.\ \citet{Held.1999.Tellus} for a recent overview), an eddy-induced diffusivity is found to have a timescale $T$ and a length scale $L$ associated with it such that 
\begin{eqnarray}
  \label{eq:dimanal:diff}
  \kappa  \propto \frac{L^2}{T} \propto \frac{f}{\sqrt{Ri}} l^2 .
\end{eqnarray}
The inverse of the instability growth rate ($f \sqrt{Ri}^{-1}$, as used by \citet{Visbeck.etal.1997}) is a natural timescale in the eddy diffusion process, but the length scale is more ambiguous. $L$ may be determined by internal dynamics (cf.\ \citet{Stone.1972}), or by the external geometry of the problem (e.g.\ the width of the baroclinic zone, cf.\ \citet{Green.1970}). Hence,
\begin{eqnarray}
  \label{eq:diff:intext}
  \kappa_{\mbox{int}} \propto \frac{\tilde{\omega}_{\mbox{max}}}{r_{h,\mbox{max}}^2} \quad \mbox{or} \quad \kappa_{\mbox{ext}}  \propto \tilde{\omega}_{\mbox{max}} L_{\mbox{ext}}^2.
\end{eqnarray}

Since a variation in $\Delta$ or $\Delta_\nu$ is expected to affect the internal length scales, the inverse of the fastest-growing wavenumber has been taken to represent the internal length scale for $\kappa_{\mbox{int}}$. Note that $L_{ext}$ is independent of $\Delta$ and $\Delta_\nu$, which implies a difference between $\kappa_{\mbox{int}}$ and $\kappa_{\mbox{ext}}$. Furthermore, because of the dependence of $\tilde{\omega}_{\mbox{max}}$ on $\Delta$, the diffusivity and hence the eddy flux will be sensitive to $\Delta$.

\begin{figure}[tp]
  \begin{center}
    (a) \includegraphics[width=0.43\textwidth]{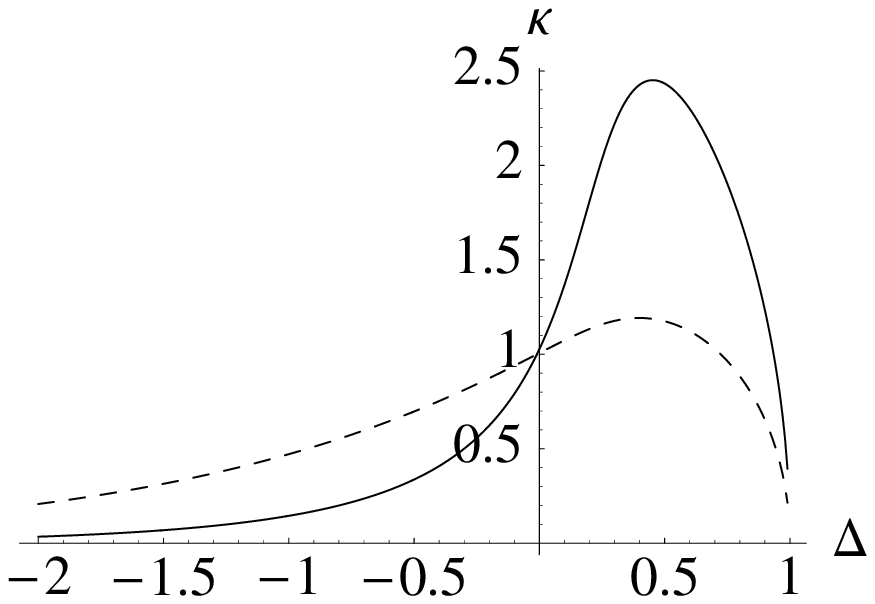}
    (b) \includegraphics[width=0.43\textwidth]{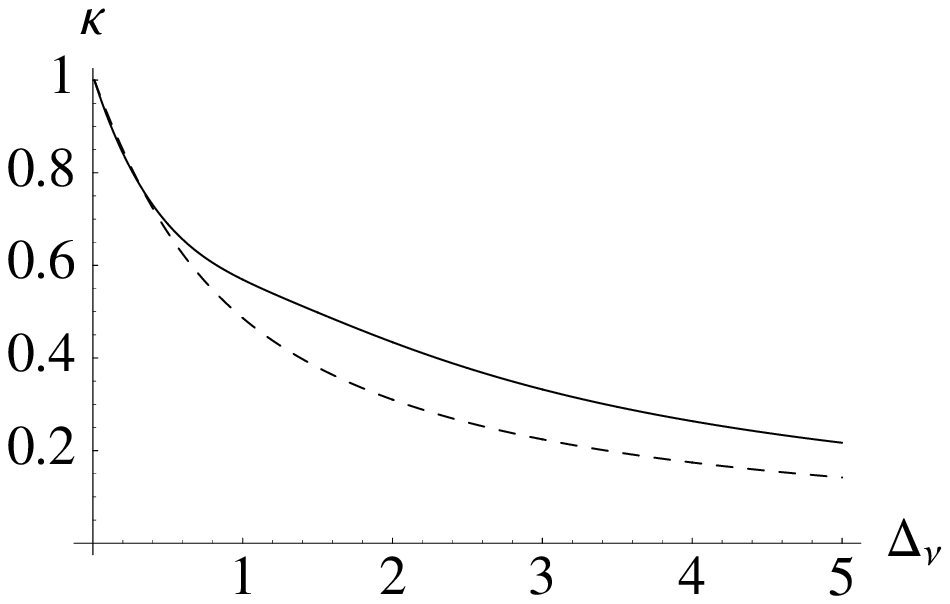}
    \caption{(a) the eddy diffusivity $\kappa_{\mbox{int}}$ (solid) and $\kappa_{\mbox{ext}}$ (dashed) for the sloping bottom (b) influence of the bottom Ekman layer on the eddy diffusivity based on the internal and external length scales, as in (a). The diffusivities corresponding to the original Eady problem are normalised to unity.}
    \label{fig:sens:eddydiff}
  \end{center}
\end{figure}

The dependence of $\kappa$ on the bottom slope and Ekman pumping ($\Delta$) is depicted in Figure \ref{fig:sens:eddydiff}(a). The largest deviations from the standard Eady problem are encountered when the eddy length scale is set by the fastest growing wave; in the diagram $\kappa$ is varying with a factor of 10 in the range $-0.5 < \Delta < 0.5$. \citet{Blumsack.Gierasch.1972} calculated the correlation $\overline{v'T'}$ as a function of $\Delta$ and obtained a result which shows a pronounced similarity to our $\kappa_{\mbox{int}}$, hereby providing support for our simple scaling argument. 

However, if the diffusivity is governed by a fixed external length scale, $\kappa$ is proportional to $ \tilde{\omega}_{\mbox{max}}$ and is therefore less sensitive to $\Delta$. In case of bottom Ekman pumping, the sensitivity of $\kappa$ to $\Delta$ is smaller than in the presence of a sloping bottom, with $\kappa$ varying only by a factor of two for $0 < \Delta_\nu  < 1$.

Hence, when the deviations from the Eady problem are not large, it appears that the bottom slope exerts a strong influence on the eddy diffusivity when the length scale is determined by internal dynamics. Furthermore, the possibility for complete stabilisation in the case of a sloping bottom (implying zero diffusivity), and the destabilisation of short waves in the Ekman layer case call for a development of new eddy parameterisations for coastal waters.

\section{Conclusions}
\label{sec:conclusions}

The baroclinic instability  mechanism requires the interaction between two propagating PV disturbances. In the Eady problem these disturbances result from the divergence of vertical velocity at the solid boundaries, and are therefore boundary-trapped. The relation between the isopycnal slope and the bottom slope (or ratio of Ekman depth to horizontal wavelength) is the single most important parameter in this problem.

The propagation speed of the disturbances influences the stability properties of the problem. An impermeable boundary which is not horizontal leads to changes in the phase speeds of the disturbances. In particular, the existence of short- and long-wave cutoffs appears to be especially sensitive to the prescribed boundary conditions via the modifications these give rise to in the boundary phase speeds.

Therefore it seems that an accurate representation of the Eady instability growth rate in an eddy parameterization would require quite detailed knowledge of the phase speeds of free Eady waves on horizontal boundaries. These phase speeds, on the other hand, are sensitive to the details of the boundary even in the quasi-geostrophic formulation. One may expect this sensitivity to be carried over to mildly non-quasi-geostrophic dynamics. 

The importance of the trapping boundary for disturbances is emphasized in this study. In the Ekman-pumped Eady problem, all amplifying modes are top-trapped and inclined against the shear in the Eady amplification sense. The effect of Ekman pumping on the long waves is to reduce the growth rate, but the same structural change causes a mild destabilisation for short waves. The pumping translates the vorticity associated with the top disturbance into a divergence in the bottom layer, thereby introducing a small disturbance at the bottom. The top disturbance and its ``slave'' on the bottom have a phase shift, which is responsible for the destabilisation. Remarkably enough, a single disturbance in combination with a suitable boundary condition suffices for the generation of instability.

The fairly large changes in the disturbance structure for bottom slope or Ekman pumping are probably a useful guide when analysing observations and model results, and since these effects are almost inevitably present in laboratory investigations, numerical models and observations, the classical Eady problem may not be the best guide when interpreting experimental results.

The solutions derived in this study indicate that idealized numerical eddy-transport experiments without bottom slope or Ekman pumping may not yield universally valid parameterizations. As has been shown here, there are regions of parameter space where the stability properties of the quasi-geostrophic problem differ markedly from those found in the basic Eady problem. Most importantly, there is a possibility for the removal of the necessary condition of instability when the isopycnals are aligned with the bottom slope. For larger slopes, Kelvin-wave dynamics become important and the stability properties may be expected to differ from those studied here.

However, for small slopes and small Ekman numbers, the maximum growth rates do not differ very much from those found in the unmodified Eady problem. This probably explains the reasonably high level of success enjoyed by eddy parameterizations based on the Eady growth rate. 

The vertical-velocity and isopycnal-slope perspective applied throughout this study serves as an interesting complement to the ``IPV thinking'' due to \citet{Hoskins.etal.1985}. The straightforwardness with which the stabilizing influence of topography can be understood within the present conceptual framework is a good argument for considering the interior as well as the boundary contributions to vortex stretching in terms of divergence of vertical velocity, as opposed to the less intuitively evident $\delta$-sheets of PV. Since boundaries have such a strong influence on the evolution of baroclinic processes, a robust specification of the boundary condition would be important for inversion purposes. In addition, the specification of a relevant boundary density distribution in the real atmosphere or ocean may be more difficult than the specification of large-scale isopycnal slope for use in the inversion process. 

The parameterisation derived by scaling arguments shows a strong decrease of eddy diffusivity with an increasing bottom slope when the mixing process is governed by an internal length scale. Such a decrease is consistent with the persistence of buoyant coastal currents. However, the quasigeostrophic theory does not appear to fully explain the observed persistence, since even with opposing isopycnal and bottom slopes, there is a narrow range of unstable wavenumbers. This unstable range would in principle be sufficient to cause a disintegration of such currents. But in fact, the unstable range is so narrow that a small modification of the boundary wave propagation properties may be capable of stabilizing the flow \emph{in toto}, thus bringing the theory into harmony with observations. Such a modification is, however, probably beyond of the scope of quasi-geostrophic analysis.

\subsection*{Acknowledgements}

Financial support from Maj and Tor Nessling Foundation and University of Helsinki funds is gratefully acknowledged. This study has its roots at the Department of Geophysics, University of Helsinki, and was considerably substantiated during an enjoyable stay at Meteorological Institute, University of Stockholm, which was financially supported by Nordiska Forskerutdanningsakademin (NorFA). Dr. J. Nilsson is deeply acknowledged for discussions related to neutral edge waves and branch cut discontinuities. Dr. B. Rudels, prof. E. Holopainen and in particular prof. P. Lundberg improved the manuscript with their comments.

\bibliography{journals,omat,baltic,turbmix,models,misc,meso,methods}
\bibliographystyle{agu}

\end{document}